\documentclass[twocolumn,twocolappendix]{aastex631} %linenumbers
\pdfoutput=1 %for arXiv submission

\accepted{October 15, 2024}

\shorttitle{Evolving MW Radial Profiles}
\shortauthors{Patel, Chatur \& Mao}

\begin{document}

\title{Temporal Evolution of the Radial Distribution of Milky Way Satellite Galaxies}

\correspondingauthor{Ekta Patel}
\email{ekta.patel@utah.edu}

\author[0000-0002-9820-1219]{Ekta~Patel}\thanks{NASA Hubble Fellow}
\affiliation{Department of Physics and Astronomy, University of Utah, 115 South 1400 East, Salt Lake City, Utah 84112, USA}

\author[0009-0000-9825-9755]{Lipika Chatur}
\affiliation{Department of Astronomy, The University of Texas at Austin, 2515 Speedway Boulevard, Austin, TX 78712, USA}

\author[0000-0002-1200-0820]{Yao-Yuan~Mao}
\affiliation{Department of Physics and Astronomy, University of Utah, 115 South 1400 East, Salt Lake City, Utah 84112, USA}

\defcitealias{patel20}{P20}
\defcitealias{samuel20}{S20}
\defcitealias{buch24}{B24}
\begin{abstract}

The Milky Way (MW) is surrounded by dozens of satellite galaxies, with six-dimensional (6D) phase space information measured for over 80\% of this population. The spatial distribution of these satellites is an essential probe of galaxy formation and for mapping the MW's underlying dark matter distribution. Using measured 6D phase space information of known MW satellites, we calculate orbital histories in a joint MW+LMC potential, including the gravitational influence of the LMC on all satellites, on the MW's center of mass, and dynamical friction owing to both galaxies, to investigate the evolution of the MW's cumulative radial profile. We conclude radial profiles become more concentrated over time when we consider the LMC's gravitational influence and the group infall of LMC-associated satellites. The MW's radial distribution is consistently more concentrated at present-day, 1 Gyr, and 2 Gyr ago compared to recent surveys of nearby MW-like systems. Compared to MW-mass hosts in cosmological, zoom-in simulations, we find the MW's radial profile is also more concentrated than those of simulated counterparts; however, some overlap exists between simulation results and our analysis of the MW's satellite distribution 2 Gyr ago, pre-LMC infall. Finally, we posit radial profiles of simulated MW-mass analogs also hosting an LMC companion are likely to evolve similarly to our results, such that the accretion of a massive satellite along with its satellites will lead to a more concentrated radial profile as the massive satellite advances toward its host galaxy.

%Previously, studies comparing observed and simulated MW-mass galaxies typically focused on their present-day characteristics, neglecting historical evolution.


\end{abstract}

\section{Introduction} 
\label{sec:intro}

It is well known that dozens of known satellite galaxies and counting surround the Milky Way (MW) and Andromeda (M31). These satellites are largely low-mass dwarf galaxies often used to test cold dark matter theory with a cosmological constant ($\Lambda$CDM) on small scales. The abundance and spatial distribution of satellites can also illuminate the properties and underlying distribution of the dark matter halos surrounding the MW and M31. 

In the era of high precision astrometry, observatories such as the \textit{Hubble Space Telescope} and \textit{Gaia} are providing precise proper motions for low mass dwarfs to the edge of the Local Group \citep[LG; e.g.,][]{vdm12ii, sohn13, sohn17, sohn20, bennet24}. These observatories will be complemented shortly by data from \textit{JWST} and the \textit{Nancy Grace Roman Space Telescope}, which will deliver additional precise proper motions for nearby galaxies.

Combined with line-of-sight (LOS) velocities and distances, proper motions yield 6-dimensional (6D) phase space information that can aid in reconstructing the recent accretion and evolutionary history of satellites in the LG. Indeed, the availability of this 6D data has already helped us make great strides in improving our knowledge of the Local Group's dynamical history. Some examples include determining the orbit of the LMC around the MW \citep{kallivayalil13}, calculating the timing of the future MW-M31 encounter \citep[e.g.,][]{vdm12,vdm19}, constraining the orbit of M33 around M31 \citep[e.g.,][]{patel17a}, studying the infall of satellites with the LMC \citep[e.g.][]{erkal20, patel20, battaglia22}, exploring the thin, planar structure of MW and M31 satellites \citep[e.g.][]{libeskind05, pawlowski19, sohn17, sohn20}; and estimating the masses of the MW, M31, and the LG as a whole \citep[e.g.][]{patel17b,patel18a, patel23, chamberlain22,benisty22, benisty24}.

Improved methods for modeling the orbits of satellite galaxies, especially the MW's most massive satellite galaxy -- the LMC -- have proven to be especially useful. Using \textit{HST} proper motions, \citet{kallivayalil13} showed the LMC is likely on first infall into the halo of the MW, having just completed a pericentric passage \citep[see also][]{besla07}. It is also now known that the total mass of the LMC is likely $1-3\times10^{11}\, M_{\odot}$, or at least 10\% the mass of the MW \citep[e.g.,][]{shipp21, watkins24}. 

With a precise mass and orbit for the LMC, several authors have made recent breakthroughs in understanding the impact of the LMC on the MW and corresponding substructures. These include quantifying the displacement and reflex motion of the MW's center of mass, predicting the formation of a dark matter wake in the MW's halo, and assessing biases in MW mass estimators resulting from the MW's response to the gravitational influence of the LMC \citep[e.g.][]{gc21a, erkal20, lilleengen22, vasiliev23, kravtsov24}. Both predictions and observations of these phenomena \citep[e.g.,][]{erkal21} have collectively indicated the MW is not in equilibrium, contrary to previous assumptions.

Given this new knowledge, we must revisit traditional ways we have attempted to place the MW in the context of MW-mass galaxies at low redshift. Recently, both the Satellites Around Galactic Analogs \citep[SAGA;][]{mao24} survey and the Exploration of Local VolumE Satellites \citep[ELVES;][]{carlsten22} surveys have completed a census of satellite galaxies around MW-mass hosts down to $M_V = -11.9$ and $M_V = -9$, respectively \citep{mao24, carlsten22}. Together, these surveys have taken a census of satellites around $\sim$130 systems, providing a statistically significant sample for comparison with satellites of the MW. 

Alongside these critical observational advances, studies using suites of cosmological zoom-in simulations of MW-mass halos have also proven exceptionally useful in placing the properties of MW satellites in context. We refer readers to \citet[e.g.][]{samuel20,samuel21, samuel22} and \citet{santistevan23, santistevan24} for analyses of the FIRE-2 cosmological zoom-in baryonic suite, Latte, centered on MW-like galaxies, and the ELVIS on FIRE suite centered on LG-like pairs. For studies of subhalos around MW-mass halos in dark-matter-only suites of cosmological zoom-in simulations, we refer readers to \citet[e.g.,][]{ kravtsov24} and \citet[][hereafter \citetalias{buch24}]{buch24} who study the Caterpillar simulations and Symphony/Milky Way-est dark matter only simulations, respectively. %This includes but is not limited to comparisons between satellite orbits, their spatial distribution, and quenching times.

Previously, studies comparing observed and simulated MW-mass galaxies typically focused on their present-day characteristics, neglecting historical evolution. With the advent of 6D phase space data and orbital modeling, we can conduct more comprehensive analyses of the MW's properties across different epochs. This advancement enables us to assess how typical the MW is among similar-mass galaxies in both the local Universe and simulations and whether its current evolutionary stage is representative of MW-mass analogs. In light of recent findings on the significant impact of the LMC on the MW over the past ~2 Gyr, it is especially crucial to also account for its impact on the evolution of MW substructures.

In this paper, we quantify how the radial distribution of MW satellite galaxies changes as a function of time, including the influence of the LMC. This paper is organized as follows. In Section \ref{sec:data}, we introduce the observational data used as input to the orbital models described in Section \ref{sec:methods}. Section \ref{sec:methods} also details how radial profiles are generated as a function of time. In Section \ref{sec:results}, we present our results for the total sample of MW satellites and when we exclude the impact of the LMC. Section \ref{sec:discussion} includes comparisons to surveys of MW-mass galaxies and the properties of simulated MW-mass analogs. Finally, we summarize and conclude in Section \ref{sec:conclusions}.

\section{Data}
\label{sec:data}
 We use 6D phase space information derived from the combination of LOS velocity, distance modulus, and proper motions to integrate the orbital histories of MW satellite galaxies. From these orbital histories, we extract distances with respect to lookback time to construct radial profiles at present and past epochs. 

Our primary observational data is from \citet{mcandvenn2020b}, which used \textit{Gaia} eDR3 to uniformly derive proper motions for 58 MW satellites. We also adopt the LOS velocities provided in \citet{mcandvenn2020b} and the R.A., Dec., and distance moduli provided in \citet{mcandvenn2020a}. Absolute magnitude estimates are from \citet{mcconnachie12} and \citet{simon19}. 

Our sample of MW satellites includes the 45 galaxies reported in \cite{mcandvenn2020b} with a measured LOS velocity and are within 300 kpc of the MW. The faintest dwarf in our sample is Draco II \citep[$M_V=-0.8^{+0.4}_{-1.00}$;][]{simon19}. We also include the LMC and SMC in our sample. When we evaluate the present-day distribution of satellite galaxies, we include the Sagittarius dwarf spheroidal (Sgr dSph) for consistency with other recent works. However, we exclude it when we evaluate the distribution of satellites at 1 and 2 Gyr ago as it has undergone significant disruption and mass loss by tides from the MW, which are not accounted for in our orbital models \citep[e.g.,][]{law05}. 

Throughout this analysis, a distance ($D$) cut of 300 kpc $(D < 300 \, \text{kpc})$ is applied consistently at each epoch where we tabulate the MW's cumulative radial profile. The total sample of satellites is $N_{\rm sat}=48$ at present-day, $N_{\rm sat}=47$ at 1 Gyr ago, and $N_{\rm sat}=45$ at 2 Gyr ago, on average, when distance uncertainties are included.

\section{Methods}
\label{sec:methods}
Here, we describe the methods used to reconstruct orbital histories for each MW satellite and how orbital information is used to construct radial profiles for MW satellite galaxies.

\subsection{Orbital Histories}
\label{subsec:orbits}
We adopt and modify the methods from \citet[][hereafter \citetalias{patel20}]{patel20} to calculate the orbital histories of MW satellite galaxies through numerical integration backward in time in a joint MW+LMC gravitational potential. Orbits are initialized with the 6D phase space information (i.e., Galactocentric 3D position and 3D velocity) described in Section \ref{sec:data}. We compute a direct orbital history (i.e., the orbit resulting directly from the transformation of measured quantities to 6D phase space vectors) for each satellite in our sample, where the given satellite, the MW, and the LMC are treated as a 3-body system. 

As in other works \citep[e.g.,][]{patel20, richstein22, bennet24}, direct orbital histories represent one orbital solution per satellite galaxy. They do not account for the measurement uncertainties on LOS velocity, distance modulus, and proper motions. We return to these uncertainties at the end of this section.

Some important modifications to the \citetalias{patel20} orbit methodology for this work include:

\begin{itemize}
    \item The MW's dark matter halo is modeled as a Hernquist sphere with a mass of $M_h=1.57\times10^{12} M_{\odot}$ and scale length $a_{\rm halo}=40.85$ kpc. This $M_h$ corresponds to a viral mass of $M_{\rm vir}=1.2\times10^{12} M_{\odot}$ and a viral radius of $R_{\rm  vir}=279$ kpc. 
    \item The MW's disk is a Miyamoto-Nagai profile with a mass of $M_d=5.78\times10^{10}\, M_{\odot}$ with disk scale length $r_a=2.4$ kpc and disk scale height $r_b=0.5$ kpc. 
    \item The MW's bulge is a Hernquist sphere with a mass of $M_b=1.4\times10^{10}\,{\odot}$ and scale length $a_{\rm bulge}=0.7$ kpc. The MW's disk, halo and bulge parameters are chosen to yield best-fit models to the observed rotation curve of the MW \citep{mcmillan2017}.
    \item The LMC is modeled as a one-component dark matter halo following a Hernquist profile \citepalias[i.e., there is no stellar disk as in][]{patel20}. The adopted virial mass is $M_{\rm vir}=1.8\times10^{11} M_{\odot}$ (the fiducial LMC mass in \citetalias{patel20}) with a Hernquist scale length $r_a=20$ kpc. The scale length is determined by finding the best fit to the LMC's measured rotation curve \citep{vdM2014}.
    \item Dynamical friction owing to the MW and the LMC are both included; however, the dynamical friction satellites experience as they pass through the LMC's halo is always in effect, whereas in \citetalias{patel20} it was only implemented when a satellite was inside the outer equal density contour of the MW-LMC potential.
\end{itemize}

These modifications are adopted to reflect the initial conditions for the MW-LMC simulations presented in \citet{gc19}\footnote{The orbit methodology used here will be the basis for a forthcoming \texttt{Python} package that will allow users to integrate orbits of MW substructures in a joint MW+LMC potential both in rigid and live halos (Garavito-Camargo et al., in prep. \& Patel et al., in prep.). Here, we adopt the rigid setup.}. In this setup, the LMC is on a first infall orbit where it starts at the MW's virial radius ($R_{\rm vir}=279$ kpc) at 2.22 Gyr ago and arrives at its observed position today after passing through pericenter 60 Myr ago \citep[see the orbit of LMC3 in Fig 2 of][]{gc19}. 

As in \citetalias{patel20}, measurement errors on LOS velocity, distance, and proper motion for each satellite are accounted for by drawing 1,000 Monte Carlo samples from the joint 1$\sigma$ uncertainties of these measured quantities. Each drawing is then converted to a 6D phase space vector in Galactocentric Cartesian coordinates and used to initialize a unique orbit per MW satellite. This process is repeated 1,000 times in the combined MW+LMC potential, yielding 1,000 possible orbital solutions for each MW satellite. In Section \ref{subsec:crp}, we will use distances extracted from these orbits to build radial profiles as a function of time.

We ignore the 6D phase space uncertainties of the LMC as these are much smaller than those of the fainter satellite galaxies in our sample. The most significant uncertainties for satellite orbits are the choice of MW and LMC masses, followed by the uncertainties on proper motions and distances. For example, keeping the adopted mass of the LMC fixed and increasing the mass of the MW can allow for multiple passages of the LMC around the MW \citep[e.g.,][]{patel17a,patel20}.

To isolate the LMC's gravitational influence, we also consider a rigid potential with just the MW with parameters identical to those used in the combined MW+LMC potential. For the rigid MW potential, we also compute 1,000 orbits for each satellite galaxy as a control sample.

Our orbit methodology, which uses a multi-component potential with a rigid halo for the MW and a rigid, spherical halo for the LMC does not account for the halo deformations the MW experiences as the LMC passes through pericenter \citep{gc19}. It also does not track the tidal mass loss experienced by the LMC and the formation of the dark matter wake trailing the LMC \citep{gc19}. In future work, we will explicitly account for these halo perturbations using gravitational potentials that capture the full time-evolution of the MW+LMC interaction simulated in \citet{gc19} by employing basis function expansions \citep{gc21a} to reconstruct the joint MW+LMC potential.

\subsection{Constructing Cumulative Radial Profiles} 
\label{subsec:crp}
Our goal is to explore the evolution of the radial distribution of MW satellites over the last $\sim$2 Gyr, equivalent to the time it takes the LMC to move from the edge of the MW's halo to its present-day position in a first infall scenario. At each time step (0, 1, 2 Gyr ago), we calculate the cumulative radial profile (also referred to as radial profile throughout) for the set of satellites within 300 kpc of the MW. 
The cumulative radial profile sums up the number of satellite galaxies within a given distance. We build 1,000 radial profiles for each time step by extracting distances from the 1,000 orbital histories computed for each satellite galaxy (see Section \ref{subsec:orbits}). 

\begin{figure*}

    \centering
    \includegraphics[width=0.49\textwidth, trim=7mm 2mm 0mm 0mm]{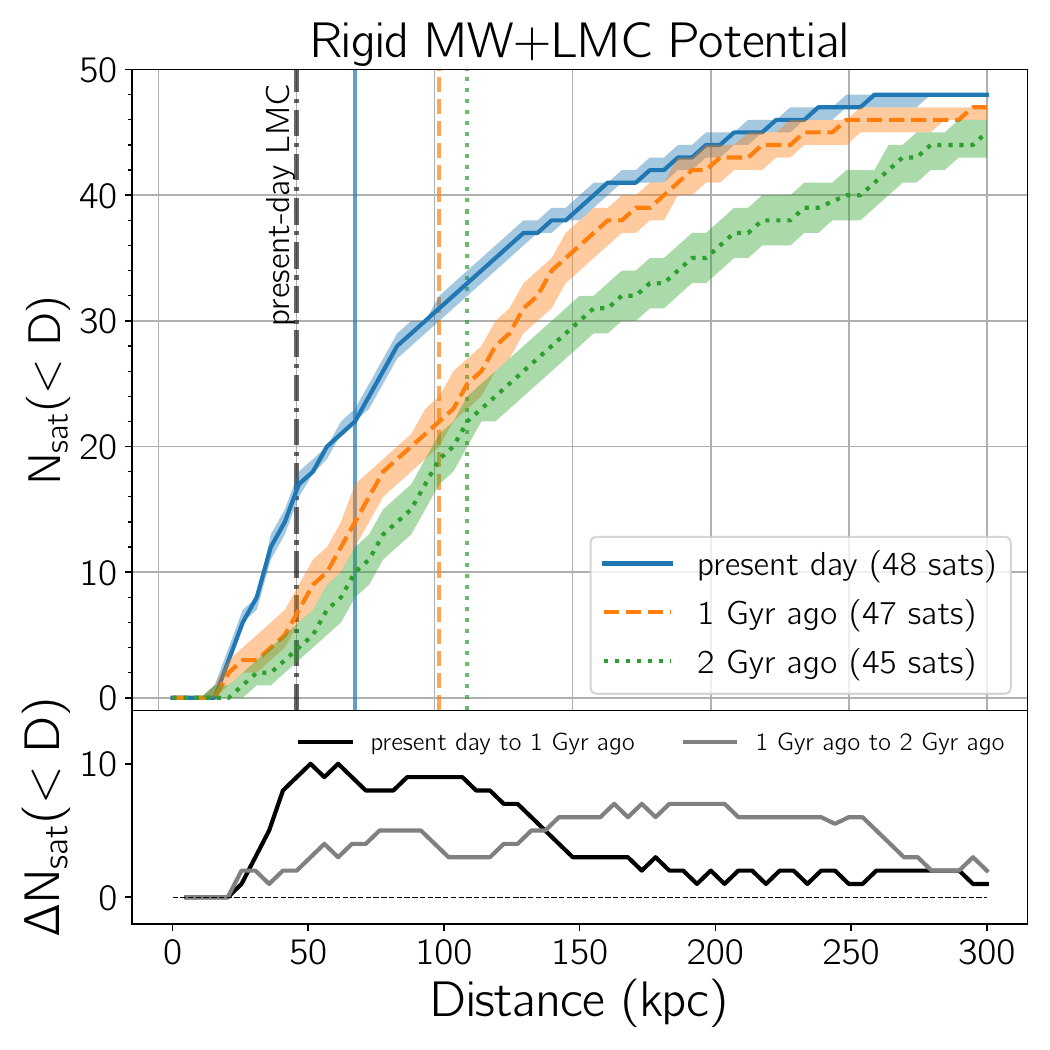} 
    \includegraphics[width=0.49\textwidth, trim=0mm 2mm 7mm 0mm]{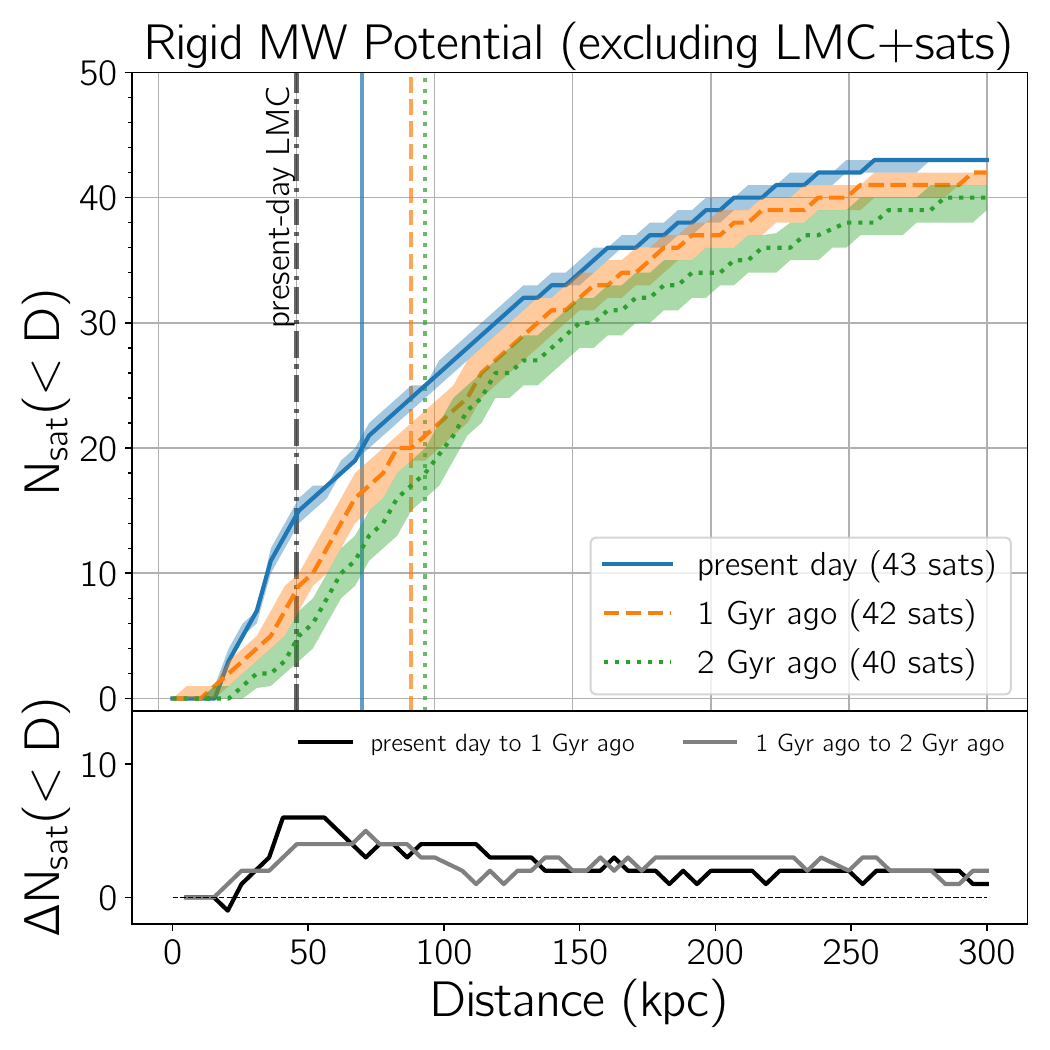} 
    \caption{\textbf{Left:} \textit{(Top)} The cumulative number of satellite galaxies as a function of 3D distance relative to the MW at present-day (solid blue), 1 Gyr ago (dashed orange), and 2 Gyr ago (dotted green) from orbits calculated in a joint MW+LMC potential. Shaded regions correspond to the 68 percent confidence intervals on the radial distribution at each time. Colored vertical lines illustrate the $R_{50}$ value for each data set, the radius at which half of the satellites are encompassed. The black vertical line shows the present-day location of the LMC. The legend indicates how many satellites are included at each time step. We do not correct for completeness. \textit{(Bottom)} The difference between pairs of radial distributions as a function of 3D distance. The black line shows the difference between the present-day and 1 Gyr ago radial profiles, while the gray line indicates the difference between the 1 and 2 Gyr ago distributions. Radial profiles are less concentrated at previous epochs compared to today. The largest differences in radial profiles between present-day and 1 Gyr ago are in the innermost regions (D $<$ 150 kpc), while the difference between the radial profile at 1 and 2 Gyr ago is more apparent in the outer regions (D $>$ 150 kpc). \textbf{Right:} \textit{(Top)} Same as left panel, but the gravitational influence of the LMC has been removed, as well as the LMC plus four satellites within $R_{\rm vir}$ of the LMC at infall (2.2 Gyr ago) and today. These satellites include the SMC, Carina III, Phoenix II, and Sculptor. \textit{(Bottom)} The difference between radial distributions for present-day and 1 Gyr ago (black) and 1 Gyr ago to 2 Gyr ago (gray). The radial distributions evolve minimally compared to Figure \ref{fig:crd_main}, implying that the presence of the LMC and its satellites drive the redistribution of satellites over time.}
    \label{fig:crd_main}
\end{figure*}

The top left panel of Figure \ref{fig:crd_main} shows the resulting radial profiles at present-day (solid blue line), 1 Gyr ago (dashed orange line), and 2 Gyr ago (dotted green line). The lines represent the medians, or 50th percentiles, of the 1,000 radial profiles generated for each time step. The shaded regions in the top panel of Figure \ref{fig:crd_main} encompass the [15.9, 84.1] percentiles (68\%) around the median radial profile. 

As distance uncertainties grow proportionally with the timescale of orbital integration, the extent of the shaded region at 2 Gyr ago is the largest, followed by the radial profile at 1 Gyr ago sample, with the present-day radial profile having the smallest uncertainty. In other words, the present-day distance uncertainties are equivalent to the measurement error on the distance modulus for each satellite, while the distance uncertainty at 1 Gyr ago and 2 Gyr ago simultaneously account for the propagation of all uncertainties on LOS velocity, distance, and proper motion.

\section{Results}
\label{sec:results}
In this section, we quantify the properties of radial profiles as a function of time and determine what drives the evolution between different epochs. We also examine the influence of the LMC on the orbits of MW satellites and, subsequently, on the radial distribution of the satellite population.

\subsection{Properties of Radial Profiles in the MW+LMC Potential}
\label{subsec:properties}

The top left panel of Figure \ref{fig:crd_main} illustrates the radial profiles for all satellites within 300 kpc of the Galaxy's center at present-day, 1 Gyr ago, and 2 Gyr ago, respectively. The total number of satellites included in the median radial profile at each time is 48 (present-day), 47 (1 Gyr ago), and 45 (2 Gyr ago). To quantify the concentration of the radial profiles at each time, we use the $R_{50}$ statistic, the radius encompassing half of the total satellite population. Since the number of satellites included in the present-day, 1 Gyr ago, and 2 Gyr ago samples is not identical, $R_{50}$ is calculated self-consistently with respect to each sample's size. We find the following values for $R_{50}$: 71 kpc (present-day), 102 kpc (1 Gyr ago), and 112 kpc (2 Gyr ago), as denoted by the vertical lines in the top panel of Figure \ref{fig:crd_main}. The present-day location of the LMC is marked with a black dot dashed line.

The present-day radial profile (blue line) is most highly concentrated compared to the radial profiles at previous epochs. While $R_{50}$=71 kpc implies 50\% of satellites reside within $D <71$ kpc, nearly 80\% of the total population is located within 150 kpc. On the other hand, the radial profiles at 1 and 2 Gyr ago have shallower slopes within 150 kpc, implying MW satellites were more spatially dispersed in the past. Direct orbital histories from Section \ref{subsec:orbits} also show that $\sim$40\% of satellite galaxies are moving toward the MW approaching present-day on first infall orbits, driving the decreasing $R_{50}$ values (and slopes) approaching present-day. Beyond 150 kpc, the radial profiles at present-day and 1 Gyr ago are qualitatively consistent, differing by just one additional satellite in the present-day sample. In comparison, the radial profile at 2 Gyr ago grows more gradually at $D>150$ kpc.

The orbits of MW satellites in the rigid MW+LMC potential also respond to the reflex motion and displacement of the MW's center of mass as the LMC passes through the MW's halo. This phenomenon shifts the MW's center of mass by up to 40 kpc over the last 2 Gyr, with 80\% of this effect taking place in just the previous Gyr \citep{gc21a}. The impact of the LMC on the MW likely explains the significant evolution of the radial profile, particularly between 1 Gyr ago and today. 

Furthermore, \citet{kravtsov24} recently analyzed the highest resolution, dark-matter-only simulations in the Caterpillar suite, which zooms in on MW-mass halos \citep{griffen16}, to investigate the impact of an LMC-mass analog on the distance and velocity distribution of MW satellites\footnote{\citet{kravtsov24} also use their results to develop a new mass estimator correcting for the impact of the LMC on satellite kinematics, finding $M_{200c} = 9.96\pm1.45\times10^{11}\, M_{\odot}$. Our assumed MW mass for satellite orbits (see Section \ref{subsec:orbits}) is consistent within the scatter of the \citet{kravtsov24} mass results.}. These authors use two specific halos from the Caterpillar suite hosting a massive satellite like the LMC that recently ($z\approx0.04-0.05$) passed through a close, pericentric passage ($r_{\rm peri}=40-50$ kpc). In Figure 6, they show the radial distribution of satellite positions becomes more concentrated at pericenter and corresponds to a high-velocity tail in the cumulative 3D velocity distribution of MW satellites owing to the response of the MW analog to the LMC analog's passage. %an example of how the radial distribution of subhalos around one of these hosts changes before ($z=0.1 \sim t_{lookback}=1.35$ Gyr), during ($z=0.05 \sim t_{lookback}=0.7$ Gyr), and after ($z=0 \sim t_{lookback}=0$ Gyr) the LMC analog is at pericenter. They conclude

In our analysis, the LMC passes through pericenter at $\sim$60 Myr, approximately corresponding to our present-day results in Figure \ref{fig:crd_main}, where we also see the highest concentration in the radial distance distribution. This agrees with the results of \citet{kravtsov24}. In Section \ref{subsec:lmcsats}, we will further examine the influence of the LMC on the distribution of MW satellites.

The bottom left panel of Figure \ref{fig:crd_main} shows the difference between the cumulative radial profiles at each pair of time steps. We denote this difference (also referred to as a differential radial profile throughout) as $\rm \Delta N_{sat}(<D)$, the difference in the cumulative number of satellites present within a specific distance at two different times. The black line represents the radial profile at 1 Gyr ago subtracted from the present-day radial profile, while the gray line shows the radial profile at 2 Gyr ago subtracted from the radial profile at 1 Gyr ago. Whether $\rm \Delta N_{sat}(<D)$ increases or decreases describes how satellites are redistributed across the MW's halo over time. By this definition, $\rm \Delta N_{sat}(<D)>0$ corresponds to a net increase in the cumulative number of satellites between time steps, while $\rm \Delta N_{sat}(<D)<0$ corresponds to a net decrease in the cumulative number of satellites.

In the bottom left panel of Figure \ref{fig:crd_main}, the most significant periods of evolution in the radial profiles between present-day and 1 Gyr ago (black line) occur between $\sim$25-150 kpc. Beyond 150 kpc, this differential is approximately constant, indicating little to no net change in the number of satellites at these distances. This does not imply that satellites are not moving to different radial distances; rather, the net change in the number of satellites changes minimally as a function of distance. The differential quantifying the change in radial profiles at 1 Gyr ago and 2 Gyr ago (gray line) shows consistent but more gradual differences across the full range of distances. 

In Appendix \ref{app:A}, we examine the distances and orbits of satellite galaxies to determine which specific satellites drive the evolution of radial profiles from one time step to the next.

\subsection{Removing the Influence of the LMC and its Satellites}
\label{subsec:lmcsats}

Several recent works have reported which of the ultra-faint dwarf galaxies discovered in the vicinity of the LMC/SMC are potential satellites of the LMC having entered the halo of the MW as a group \citep[e.g.,][]{jethwa16, sales17, erkal20, patel20}.
Independent methodologies based on both position and phase space arguments have been applied to conclude which satellites belong to the LMC \cite[see][for a review]{vasiliev23}. Typically, the SMC is implicitly accounted for in a first infall scenario.

\begin{figure}[t]
    \centering
    \includegraphics[width=0.5\textwidth, trim=10mm 5mm 0mm 0mm]{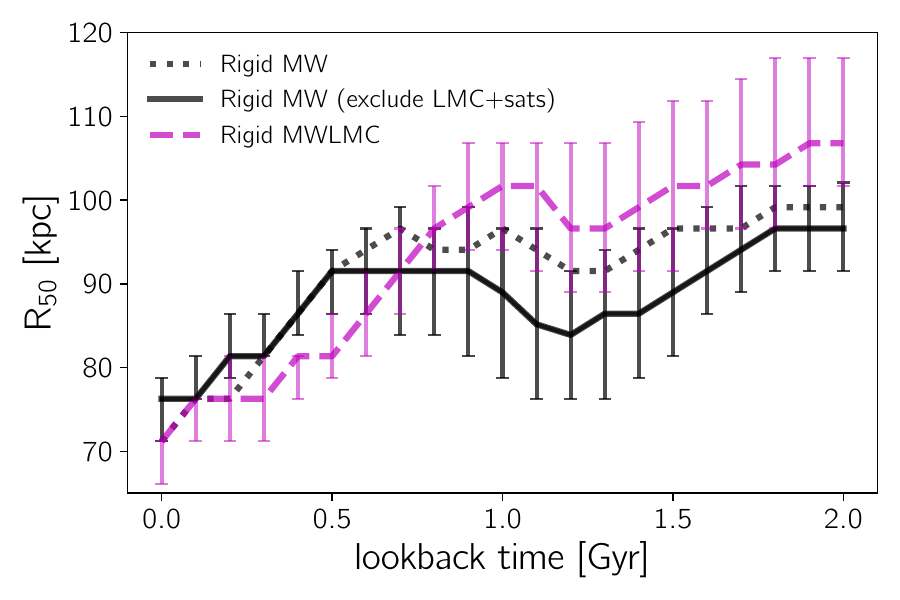}
    \caption{The evolution of $R_{50}$ with time for all satellites in the Rigid MW+LMC potential (dashed magenta line), all satellites in the Rigid MW potential (dotted black line), and satellites in the Rigid MW potential excluding the LMC and its satellites (solid black line). Lines track the median of $R_{50}$ across 1,000 radial profiles constructed at each time step. Error bars represent the 25th and 75th percentiles of $R_{50}$.}
    \label{fig:r50_time}
\end{figure}

For this analysis, we aim to isolate the influence of the LMC itself and the group of satellites that may have entered the MW's halo alongside it. We identify LMC satellites as those within $R_{\rm vir}$ of the LMC (148 kpc) and 300 kpc of the MW at infall (2.2 Gyr ago) as in \citet{nadler20}. These satellites include the SMC, Carina III, Phoenix II, and Sculptor. We will revisit which MW satellites evidence a shared orbital history with the LMC using the metrics developed in \citetalias{patel20} in forthcoming work that will explore the effects of the time-evolving joint MW+LMC potential on MW satellites (Garavito-Camargo et al., in prep.; Patel et al., in prep.). 

We repeat the exercise of building MW radial profiles by i.) excluding the LMC's gravitational influence (i.e., using orbits computed in a rigid MW-only potential) and ii.) removing the SMC, Carina III, Phoenix II, and Sculptor from the satellite sample at all time steps. The top right panel of Figure \ref{fig:crd_main} presents the resulting radial profiles. The total number of satellites in each sample is as follows: 43 (present-day), 42 (1 Gyr ago), and 40 (2 Gyr ago). The corresponding $R_{50}$ values are 74 kpc (present-day), 92 kpc (1 Gyr ago), and 97 kpc (2 Gyr ago). 

Compared to the results presented for the rigid MW+LMC radial profiles in Section \ref{subsec:properties} (top left panel of Figure \ref{fig:crd_main}), the top right panel of Figure \ref{fig:crd_main} qualitatively shows less evolution. This is quantitatively captured by the $R_{50}$ markers in the top right panel of Figure \ref{fig:crd_main}, which have a significantly narrower range, approximately half the span of the $R_{50}$ values presented in the top left panel of Figure \ref{fig:crd_main}. The bottom right panel of Figure \ref{fig:crd_main} shows the difference in pairs of radial profiles, which remain approximately constant at $D>75$ kpc. Appendix \ref{app:A} further explores the redistribution of satellites with time. 

To determine the magnitude of the LMC's gravitational influence, we calculate $R_{50}$ as a function of time for \emph{all satellites} within 300 kpc of the MW in the rigid MW-only potential compared to the MW+LMC potential. The results are presented in Figure \ref{fig:r50_time}, where the black dotted line corresponds to the rigid MW results, and the dashed magenta line corresponds to the rigid MW+LMC results. While both lines intersect at present, $R_{50}$ varies across a larger range ($\sim$70-105 kpc vs. $\sim$70-95 kpc) when the influence of the LMC is captured. Figure \ref{fig:r50_time} also shows the results for the Rigid MW potential when the LMC and its satellites are removed (solid black line), as in the top right panel of Figure \ref{fig:crd_main}. The range of $R_{50}$ narrows for the rigid MW potential when these five satellites are excluded. Figure \ref{fig:crd_main} concisely illustrates the individual effects of the LMC and its satellites on the concentration of radial profiles, with the impact of the LMC being slightly more than that of its satellites.

We conclude the LMC's gravitational influence and group infall play key roles in the evolution of the MW's radial profile. Additionally, we have only considered an LMC mass of $1.8 \times 10^{11}\, M_{\odot}$, but current upper limits for the mass of the LMC reach $\sim 3.5 \times 10^{11}\, M_{\odot}$ \citep[see][for a compilation]{watkins24}. \citetalias{patel20} and \citet{gc21a} have shown the impact of the LMC on satellite orbits, and the MW increases proportionally with mass; therefore, the results presented here are expected to scale up in the case of a more massive LMC. In the absence of a massive satellite galaxy on recent infall, we expect minimal evolution in radial profiles as a function of time. We caution studies aiming to compare the radial profile of the actual MW with MW analogs in both simulations and observations to proceed with this in mind.

There are other effects to note that also contribute to the evolution of the MW's radial profile, though to a lesser degree. These effects include the changing orbital configuration of satellites (i.e., whether satellites are at pericenter, apocenter, or somewhere in between at the three epochs of interest) and satellite distance uncertainties. When distance uncertainties are ignored, radial profiles are not equivalent to those shown in Figure \ref{fig:crd_main}. This reflects that direct orbits differ from the median of 1,000 orbital solutions computed per satellite and that distance uncertainties can give rise to some evolution in the radial profiles. Finally, we remind readers that the census of MW satellites is not yet complete and specifically that a majority of MW satellites are currently closer to pericenter than apocenter \citep[see][]{fritz18}. This implies many more distant MW dwarfs are yet to be discovered. If satellites exist at larger distances, the present-day radial profiles may not exhibit such high concentrations; however, it is unclear exactly how radial profiles would evolve backward in time with a complete census of MW satellites.

\section{Discussion}
\label{sec:discussion}
Here, we compare the results reported in Section \ref{sec:results} to observations of MW-like galaxies in the local Universe and simulated MW analog galaxies. We aim to determine whether the present-day MW is more representative of its galactic counterparts, or if the MW was more representative of other MW-like galaxies at earlier epochs in its evolution.

\subsection{Comparisons to Observational Analogs of the Milky Way}
\begin{figure*}[ht]
    \centering
    \includegraphics[width=\textwidth,trim=5mm 5mm 5mm 0mm]{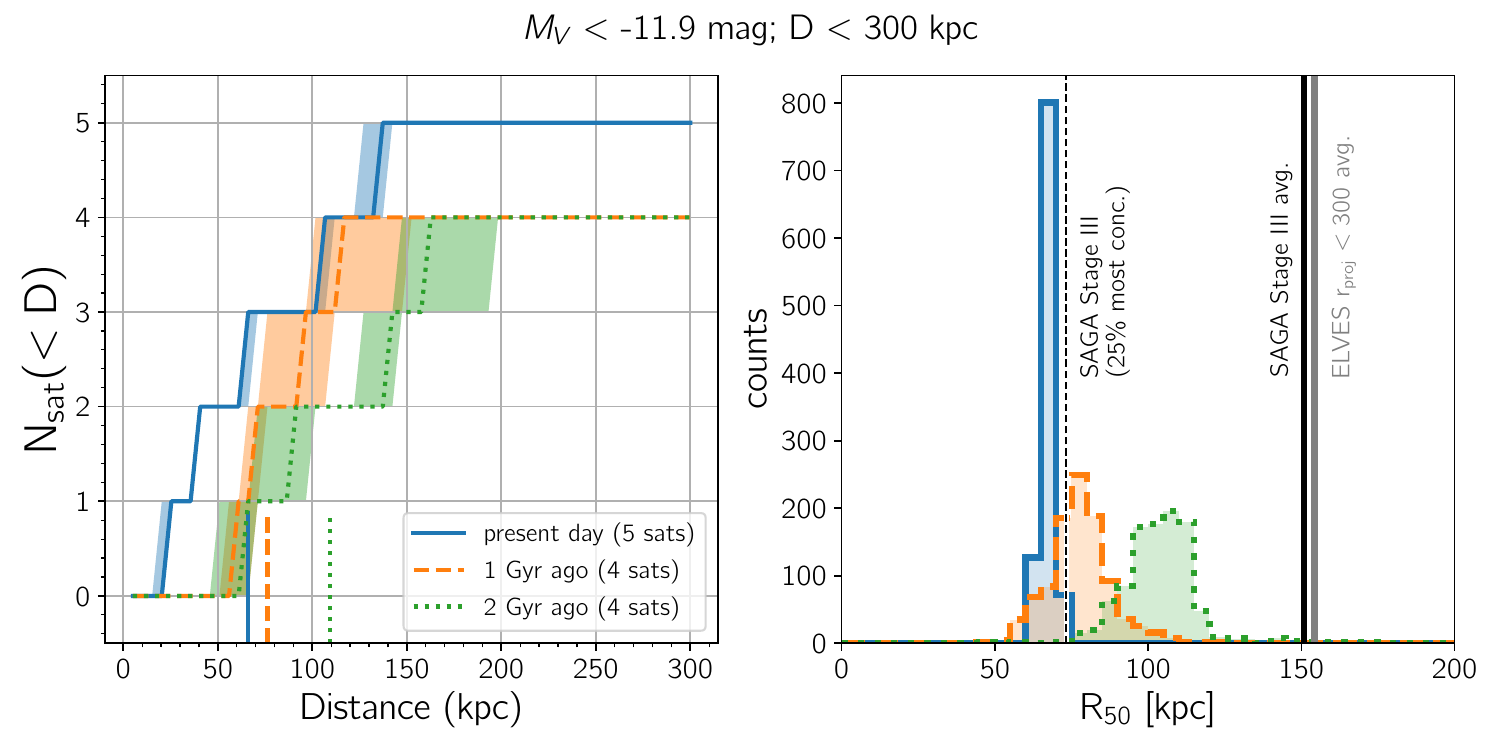}
    \caption{\textbf{Left:} The cumulative radial profile for satellites with $M_V < -11.9$ mag at $D < 300$ kpc in the joint MW+LMC potential. This magnitude selection reduces the MW sample to only four or five satellites, as indicated in the legend. \textbf{Right:} The distribution of $R_{50}$, the radius encompassing half of the satellites within 300 kpc of the MW, at present-day (blue histogram), 1 Gyr ago (orange histogram), and 2 Gyr ago (green histogram) for satellites with $M_V < -11.9$ mag. The thick black line represents the average $R_{50}$ for the 101 hosts included in Data Release 3 (DR3) of the SAGA survey \citep{mao24}. The thin, dashed black line represents $R_{50}$ for the subset of the 25\% most concentrated SAGA DR3 hosts. The thick gray line illustrates the average $R_{50}$ from a sample of hosts included in the ELVES survey \citep{carlsten22}. The MW's present-day $R_{50}$ is most consistent with the highly concentrated hosts in SAGA DR3, while the average SAGA DR3 and ELVES $R_{50}$ values are significantly larger than the MW population even at 1-2 Gyr ago.}
    \label{fig:SAGA_comparison}
\end{figure*}

Many recent efforts have been made to characterize satellite populations around MW/M31-mass galaxies beyond the Local Group \citep[e.g.,][]{spencer14,danieli17,smercina18,kondapally18, bennet19, crnojevic19, bennet20}. A diverse range of host galaxy systems have been studied to different limiting magnitudes, revealing that they exhibit a wide variety in the number of satellite galaxies they harbor and in the properties of those satellite galaxies.

We focus specifically on the results of the SAGA survey \citep{geha17, mao21, mao24} and the ELVES survey \citep{carlsten22}. The latest SAGA data release (Data Release 3, DR3) contains 101 systems at 25-40 Mpc, with satellites down to $M_V = -11.9$ mag. ELVES includes 28 systems within 12 Mpc, reaching satellites down to $M_V = -9$ mag. In what follows, we will compare the average radial profiles of host systems in these surveys to our results using the $R_{50}$ statistic. We will leave a more in-depth analysis linking the properties of satellites (i.e., color, stellar mass, quenching time) and their evolving radial distances to future work.

As the SAGA survey is only complete to a limiting magnitude of $M_{r,0} < -12.3$ mag (or $M_V < - 11.9$ mag), in this exercise, we consider only the MW satellites passing this magnitude selection\footnote{The ELVES survey is complete down to $M_V \sim -9$ mag, but we only include satellites down to $M_V = -11.9$ for consistent comparisons between SAGA, ELVES, and our results.}. This consists of the LMC, SMC, Sgr, Fornax, and Leo I. Using the distances extracted from the 1,000 orbital histories computed for each satellite in Section \ref{subsec:orbits}, we compute 1,000 radial profiles at each time step for this subsample. The resulting median radial profiles and 68\% confidence intervals are presented in the left panel of Figure \ref{fig:SAGA_comparison}. As in Section \ref{sec:results}, we consistently apply a distance cut of 300 kpc. Even with a small sample of only the brightest satellites, the evolution of the radial profile is still evident. 

The right panel of Figure \ref{fig:SAGA_comparison} shows the distribution of $R_{50}$ for the present-day (blue), 1 Gyr ago (orange), and 2 Gyr ago (green) radial profiles. Each histogram represents 1,000 $R_{50}$ values corresponding to 1,000 radial profiles. The two black vertical lines correspond to $R_{50}$ for SAGA DR3, where the thick black line represents the average of all 101 systems and the thin, dashed black line is the average of the 25\% most concentrated systems from \citet{mao24}. 

The gray line corresponds to the average $R_{50}$ from the ELVES survey \citep{carlsten22}. For ELVES, $R_{50}$ is computed for only those satellites with $M_V < -12$ mag in the ten hosts passing the SAGA host selection criteria and have coverage to 300 kpc. These ten hosts include NGC 253 (2 satellites), 628 (11), 1291 (9), 2683 (2), 2903 (4), 3115 (7), 4736 (1), 5055 (6), 5236 (7), and 5457 (5), where the numbers in parentheses indicate the number of satellites with $M_V < -12$ mag.

Generally, our $R_{50}$ values are significantly smaller at all epochs compared to the SAGA DR3 and ELVES $R_{50}$ lines at $\sim$150 kpc. The latter minimally intersects with the high tail end of our $R_{50}$ histogram at 2 Gyr ago (green histogram). On the other hand, the most highly concentrated SAGA systems are approximately consistent with our present-day MW results (blue histogram), however, the most concentrated systems in SAGA are not strictly those that host LMC-mass analogs. We conclude that MW is amongst the most radially concentrated satellite systems across observed MW analogs, even at 2 Gyr ago.

\subsection{Comparisons to Simulated Analogs of the Milky Way}
\label{subsec:sims}
Multiple recent studies have characterized the radial profile of MW analogs in both isolated and Local Group-like (LG-like) environments using cosmological zoom-in simulations. We use these studies to place our results in the context of statistically significant samples of MW analogs with a broad range of formation and evolutionary histories.

\begin{figure*}[ht]
    \centering
    \includegraphics[width=\textwidth,trim=5mm 5mm 5mm 0mm]{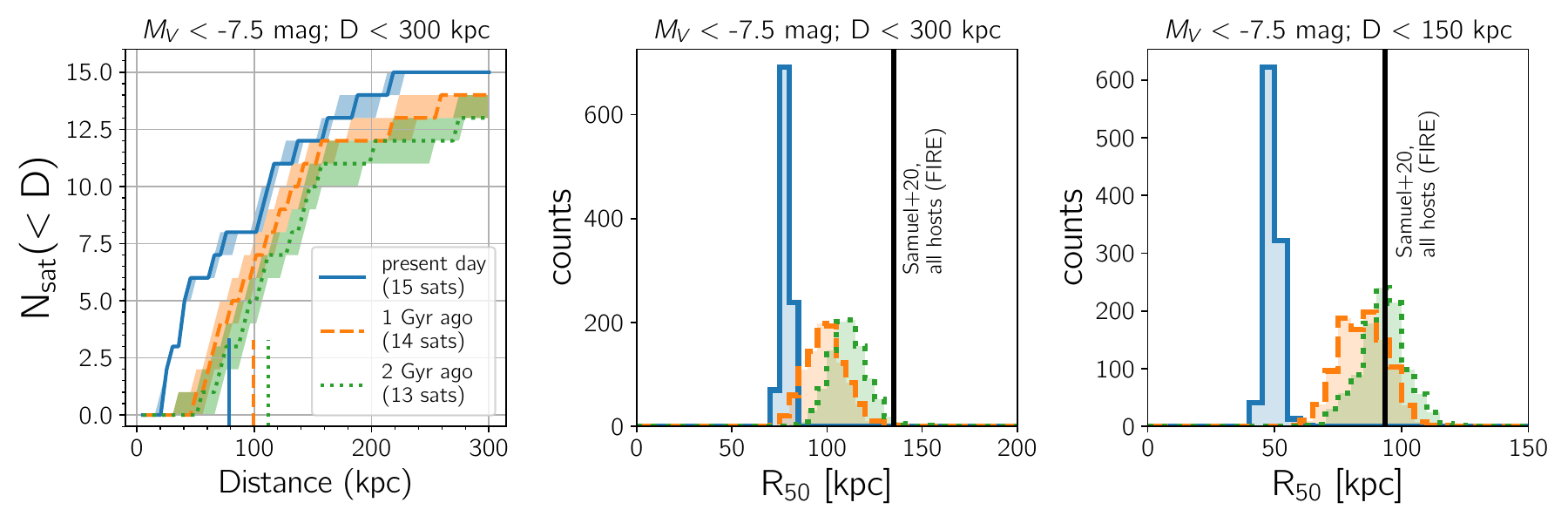}
    \caption{\textbf{Left:} Same as Figure \ref{fig:SAGA_comparison} including all satellites with $M_V < -7.5$ mag, the magnitude equivalent to the resolution of the Latte simulations. This magnitude selection limits the MW satellite sample to 13-15 satellites, as indicated by the parenthetical in the legend. \textbf{Middle:} Same as Figure \ref{fig:SAGA_comparison} for all satellites with $M_V < -7.5$ mag. The thick black line represents the average $R_{50}$ for all Latte MW analogs analyzed in \citetalias{samuel20}. Our results exhibit $R_{50}$ values below those of Latte MW analogs, with only minimal overlap between our 2 Gyr ago results. \textbf{Right:} Same as the middle panel but only including satellites within 150 kpc. The black vertical line represents the average $R_{50}$ for all Latte MW analogs analyzed in \citetalias{samuel20}, also within 150 kpc. Our results for the MW at previous epochs are in good agreement with those from Latte, suggesting the MW may have been more representative of simulated analogs in past evolutionary phases. }
    \label{fig:FIRE_comparison}
\end{figure*}

\subsubsection{Latte and ELVIS on FIRE Simulations}
\citet[][hereafter \citetalias{samuel20}]{samuel20} analyzed the radial profile of simulated MW analogs in both isolated and LG-like (i.e., with an M31-mass companion) environments. Two suites of cosmological zoom-in baryonic simulations were used, namely the Latte suite of ten halos with $M_{200m} = 0.9-2 \times 10^{12}\, M_{\odot}$ and the ELVIS on FIRE suite \citep{gk19a} which includes two pairs of MW-M31 analogs (i.e., Local Group-like pairs, also denoted as LG-like pairs). All satellites with stellar masses $M_* > 10^5 \, M_{\odot}$ and distances within 300 kpc of their host halos were considered. Results were averaged over simulation snapshots corresponding to the last $\sim1.3$ Gyr. 

\citetalias{samuel20} conclude the radial profiles of their isolated MW analogs from Latte at $z\approx0$ are broadly consistent with observations of the MW and M31 within 150 kpc. However, they predict the MW may have 2-10 galaxies with $M_* > 10^5 \, M_{\odot}$ that have yet to be discovered. Searches of the Sloan Digital Sky Survey, on the other hand, claim the MW's satellite population is complete down to $M_V = -6.5$ mag \citep[][and references therein]{simon19}. Therefore, we directly compare our results to the simulation results without an additional completeness correction.

For consistency with \citetalias{samuel20}, we select all MW satellites within 300 kpc with $M_V < - 7.5$ mag (equivalent to $M_* > 10^5 \, M_{\odot}$). This magnitude selection limits the MW satellite sample to between 13 and 15 satellites, while the Latte suite has 17 satellites within 300 kpc, on average. The left panel of Figure \ref{fig:FIRE_comparison} shows the resulting radial profiles at each time step. Again, we see evolution in the radial profiles, as shown in Section \ref{sec:results}.

The middle panel of Figure \ref{fig:FIRE_comparison} shows the distribution of $R_{50}$ for the sample of satellites with $M_V < - 7.5$. The results of all hosts in \citetalias{samuel20} are illustrated with a black vertical line. Compared to our distributions of $R_{50}$, all hosts from \citetalias{samuel20} typically have larger $R_{50}$ values, which may be in part due to the prediction of several undiscovered satellites in the outskirts of the MW ($D>150$ kpc). However, there is a narrow area of parameter space where the distribution of our results for MW satellites 2 Gyr ago (green histogram) intersects with the \citetalias{samuel20} results from the Latte suite. If we limit our satellite selection criteria only to include those within 150 kpc of the MW where \citetalias{samuel20} claim the MW population is complete, the simulated results (also limited to 150 kpc) intersect with the peak of the MW results at 2 Gyr ago, and the upper end of the MW results at 1 Gyr ago, as illustrated in the right panel of Figure \ref{fig:FIRE_comparison}.

While the Latte sample of \citetalias{samuel20} does include two host halos with an LMC-mass analog, these analogs have orbits different from the LMC orbit adopted in this work (i.e., none of them pass through pericenter during the last 1.3 Gyr, the period over which their results are averaged). Therefore, it is not straightforward to evaluate how the influence of massive satellites like the LMC might affect the results of the simulated MW analogs in Latte, on average, compared to the actual MW-LMC system. Nevertheless, we conclude that the MW was more representative of MW analogs in simulations in the past than it is now.

\subsubsection{Milky Way-est and Symphony Simulations}
\label{subsub:mwest}
More recently, \citetalias{buch24} characterized the properties of subhalos around 33 MW analogs in the ``Milky Way-est" dark-matter-only (DMO), cosmological, zoom-in simulations. These halos are specifically chosen to host LMC analogs (based on its mass ratio to the MW and its orbit) and the ancient Gaia-Sausage-Enceladus merger \citep{helmi18, belokurov18}. 
\citetalias{buch24} also study the properties of 45 isolated MW halos from the Symphony Milky Way suite as a control sample \citep{mao15,nadler23b}. Since the MW satellite sample used in this analysis is not complete to the equivalent magnitude of the MW-est resolution limits \footnote{The dark matter particle mass in the highest resolution regions of MW-est have a mass $ m_{dm} = 4 \times 10^5 \,M_{\odot}$. \citetalias{buch24} consider subhalos with present-day virial masses $M_{sub} > 1.2 \times 10^8 \, M_{\odot}$.}, we refrain from a quantitative comparison with \citetalias{buch24} and instead focus on general trends between our results.

\citetalias{buch24} find MW-est halos have $\approx 22$\% more subhalos than the Symphony hosts due to the presence of an LMC analog and its associated subhalos, identified as those subhalos within $R_{\rm vir}$ of the corresponding LMC analog. They find half (11\%) of these satellites are within $R_{\rm vir}$ of the LMC while the other half are within $2R_{\rm vir}$ at the time when this region is entirely outside of the MW analog's halo. Returning to our results from Section \ref{sec:results}, we find 4/45 (9\%) satellites are within $R_{\rm vir}$ of the LMC at infall ($\sim$ 2 Gyr ago). This validates our conclusion that the group infall of LMC satellites is one of the main drivers of our evolving radial profiles. However, DMO simulations, such as MW-est, do not account for the enhanced disruption of subhalos known to occur when a gravitational disk component is included \citep[e.g.,][]{gk17b,kelley19, wang24}. This effect is expected to reduce the number of subhalos that survive to present-day, and therefore the \citetalias{buch24} estimates should be considered an upper limit.

Based on the results presented in Figure \ref{fig:crd_main}, we predict the radial profiles of the isolated MW Symphony sample used in \citetalias{buch24} are likely to remain approximately the same if they are traced back to 2 Gyr ago in the simulation, whereas the average radial profile of the MW-est host sample is likely to evolve similarly to our Figure \ref{fig:crd_main} due to the infall of the LMC analogs and their associated satellites. This is beyond the scope of this analysis but will be the subject of future work.

\section{Summary and Conclusions}
\label{sec:conclusions}

We have quantified the MW's cumulative radial profile evolution using all known MW satellites with measured 6D phase space within 300 kpc. We traced this evolution over the last 2 Gyr, corresponding to the LMC's infall time (and its associated satellites). For comparison, we constructed radial profiles in both a joint MW+LMC potential and a rigid MW-only potential. Finally, we compared our results to observational surveys of MW-mass galaxies and cosmological zoom-in simulations of MW-mass analogs. Our main findings include:

\begin{enumerate}

    \item We find the cumulative radial profile of MW satellites becomes more concentrated over the last 2 Gyr in a combined MW+LMC potential (left panels, Fig.~\ref{fig:crd_main}), as exhibited by decreasing  $R_{50}$ values with time: 112 kpc (2 Gyr ago), 102 kpc (1 Gyr ago), and 71 kpc (present-day).  
 
    \item When we repeat our analysis in a MW-only potential removing the LMC and its four satellites (SMC, Carina III, Phoenix II, Sculptor), the evolution of the MW's radial profile is weaker (right panels, Fig.~\ref{fig:crd_main}). The $R_{50}$ values for radial profiles excluding LMC satellites are 74 kpc, 92 kpc, and 97 kpc. This range is only half that of the $R_{50}$ values for radial profiles computed in the MW+LMC potential.

    \item We attribute the increased concentration over time to the LMC's gravitational influence and the group infall of satellite galaxies associated with the LMC, though other less significant effects should also be considered. Our conclusions agree with recent work examining the radial profiles of simulated MW-mass analogs with and without an analog of the LMC's mass and orbit (see Section \ref{subsec:sims}).
    
    \item We use a subset of our satellite sample (selected based on $M_V$) to compute and compare our radial profile results to those presented in the SAGA and ELVES surveys (Fig.~\ref{fig:SAGA_comparison}). We conclude the MW's current evolutionary state is rare amongst MW-mass systems in the local Universe. Instead, the MW's present-day radial profile is most well-aligned with the 25\% most concentrated SAGA systems.
    
    \item Using a subset of galaxies (selected based on simulation resolution limits), we compute and compare our radial profiles to those of MW-mass analogs in cosmological zoom-in simulations (Fig. \ref{fig:FIRE_comparison}). Considering satellites within 300 kpc of the MW, our results show the MW is not representative of simulated MW analogs at present-day, 1 Gyr ago, or 2 Gyr ago. When we limit the sample to only those satellites within 150 kpc, we find good agreement between the simulated analogs and our results for the MW at 1 and 2 Gyr ago, implying the MW was more similar to simulated counterparts in its previous evolutionary phases.
    
    \item Based on our conclusions, we posit radial profiles of MW+LMC analogs (where analogous systems align with the estimated mass and orbit of the LMC) in simulations are expected to evolve to a greater degree compared to MW-like systems without a massive satellite companion. 

\end{enumerate} 

In forthcoming work, we will quantify the gravitational influence of the LMC and its impact on the orbits of satellite galaxies using a joint MW+LMC potential \citep[derived from the simulations of][]{gc19} accounting for the evolution of the MW and the LMC's dark matter halos over the last $\sim$2 Gyr (Patel et al., in prep.). We will re-examine whether the LMC satellites previously concluded to be associated \citepalias[e.g.,][]{patel20} holds and how this might impact the evolution of the MW's radial profile presented in this work. Furthermore, we will account for the impact of M31's gravitational influence to determine if it has any effect on the distribution of MW satellites.

It is worth noting that the observational completeness of known MW satellite population is not homogeneous, as these dwarfs were discovered using instruments with varying sensitivity, and the sensitivity also depends on angular position. This poses an interesting question of whether the the radial distribution evolves differently when these selection effects are taken into account. This will be explored in future work, but our current analysis with a more observationally complete sample (bright satellites within 150 kpc; as shown in the rightmost panel of Figure \ref{fig:FIRE_comparison}) suggests that the radial distribution still evolves inward with time.

As more Milky Way (MW) satellite galaxies are discovered, and their complete 6D phase space information is measured, it is crucial we continue to revisit our understanding of the distribution of these satellites around the MW. More than a dozen known satellites lack LOS velocities and/or proper motion data, which will significantly expand the viable sample of MW satellites once these measurements become available. These satellites are distributed across the MW's halo from 25-250 kpc. Thus, they are unlikely to change the results presented here. 

However, we have not yet considered the disrupted dwarf galaxies observed as stellar streams today. These dwarf progenitors would have been considered satellites in the past, depending on their disruption timescales. Since stellar streams tend to be within 50 kpc of the MW, including these dwarf progenitors in our radial profile analysis would likely skew the radial profiles toward lower $R_{50}$ values compared to those in this work.

When sufficient phase space information becomes available for the satellite galaxies around Andromeda (M31), we can similarly examine how its radial profile evolves. This may reveal valuable clues about M31's recent accretion history, potentially providing new insights into the formation and evolution of massive galaxy halos.

\begin{acknowledgements}  
EP is supported by NASA through Hubble Fellowship grant \# HST-HF2-51540.001-A, awarded by the Space Telescope Science Institute (STScI). STScI is operated by the Association of Universities for Research in Astronomy, Incorporated, under NASA contract NAS5-26555. EP thanks Dan Weisz for helpful discussions that contributed to the genesis of this project and Emily Strickland for providing early data files used to develop the comparison to the SAGA survey. The authors also thank Paul Bennet, Jenna Samuel, and Ethan Nadler for feedback that helped improve the quality of this manuscript. EP and LC express their gratitude to the organizers of the January 2023 Conference for Undergraduate Women in Physics at Texas Christian University. It was through this event that they were introduced to each other.
\end{acknowledgements}

\software{Numpy \citep{numpy},
  SciPy \citep{SciPy-NMeth},
  Matplotlib \citep{matplotlib},
  IPython \citep{ipython},
  Jupyter \citep{jupyter}, 
  Astropy \citep{astropy:2013, astropy:2018, astropy:2022}, 
  Gala \citep{gala}}

%\clearpage
\appendix

\section{The Radial Redistribution of MW Satellites}
\label{app:A}
\begin{figure*}[ht]
    \centering
    \includegraphics[width=0.49\textwidth, trim=5mm 0mm 0mm 0mm]{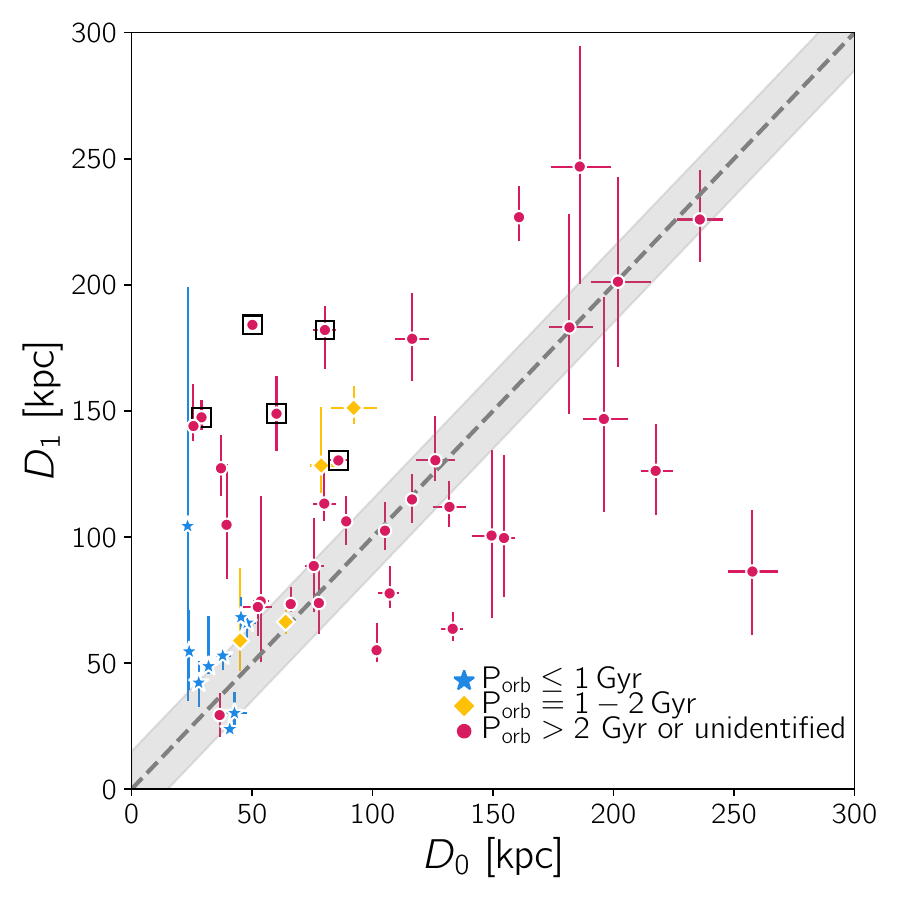}
        \includegraphics[width=0.49\textwidth, trim=5mm 0mm 0mm -2mm]{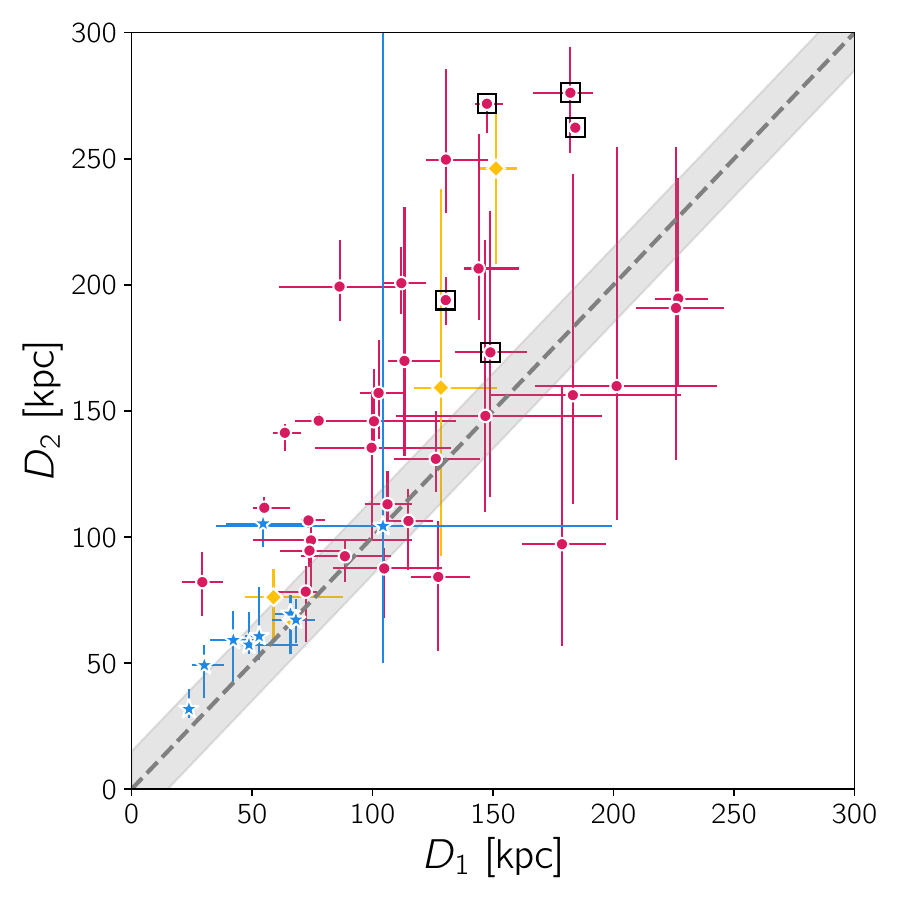}
    \caption{\textbf{Left:} The median distances of all galaxies within 300 kpc of the MW at present-day ($D_0$) and 1 Gyr ago ($D_1$). Error bars illustrate the extents of the 25th and 75th percentiles of distance from the 1,000 orbits using the Rigid MW+LMC potential described in Section \ref{subsec:orbits}. Blue stars represent satellites with short orbital periods ($\rm P_{orb} =\leq$ 1 Gyr), yellow diamonds represent satellites with intermediate orbital periods ($\rm P_{orb}=1-2$ Gyr), and pink circles represent satellites with long ($\rm P_{orb} > 2$ Gyr) or unidentified orbital periods. The gray dashed line represents a one-to-one correlation, and the corresponding shaded region encompasses $\pm$15 kpc. Open black squares denote the LMC and the four satellites identified as LMC satellites in Section \ref{subsec:lmcsats}. \textbf{Right:} Same as the left panel but at 1 Gyr ago ($D_1$) and 2 Gyr ago ($D_2$).}
    \label{fig:D1D0D2}
\end{figure*}

The bottom left panel of Figure \ref{fig:crd_main} shows significant changes in the differential radial profiles, $\rm \Delta N_{sat}(<D)$, between 1 Gyr ago and present-day, specifically at $D\sim25-150$ kpc. Using satellite distances extracted from 1,000 orbital histories computed in the rigid MW+LMC potential, determine how satellites are redistributed by visualizing distances compared to the length of each satellite's orbital period.

The left panel of Figure \ref{fig:D1D0D2} shows the median distance of all satellites at present-day ($D_0$) compared to their distances at 1 Gyr ago ($D_1$). Points are colored by the length of the orbital period ($\rm P_{orb}$) such that blue stars are satellites with $\rm P_{orb}\leq 1$ Gyr, yellow diamonds have $\rm P_{orb}=1-2$ Gyr and pink circles have either $\rm P_{orb}>2$ Gyr or an orbital period that cannot be determined during the 2 Gyr integration period. The gray dashed line represents a one-to-one correlation, and the corresponding shaded region encompasses $\pm$15 kpc. Black squares encompass the LMC and its four associated satellites.

Comparing $D_1$ and $D_0$, 30\% of satellites fall in the gray-shaded region, where distances have not changed by more than $\pm$15 kpc. In the distance range of interest, $D \sim 25-150$, most satellites with short orbital periods (blue stars) remain within $D \lesssim$ 50 kpc at both epochs. Satellites with intermediate-length orbital periods (1-2 Gyr) stay at about the same distance or move towards the MW approaching the present day. However, many satellites with long or unidentified orbital periods (pink circles) redistribute by tens of kiloparsecs in the last billion years. Nearly twice as many of these satellites fall above the one-to-one relation at $D\sim25-150$ kpc than below it in the same distance range. The former are the satellites moving towards the MW, which includes the LMC and its companion satellites, while the latter are those moving away from the MW.

The changes between radial profiles at 2 Gyr ago and 1 Gyr ago (gray line in the bottom panel of Figure \ref{fig:crd_main}) are more gradual but persist across nearly the entire 300 kpc distance range. The right panel of Figure \ref{fig:D1D0D2} shows the distances of all satellites at 1 Gyr ago ($D_1$) versus 2 Gyr ago ($D_2$), classified by the length of the orbital period.

Comparing $D_2$ and $D_1$, 28\% of satellites are in the gray-shaded region, where distances have not changed by more than $\pm$15 kpc. These are primarily satellites with short or intermediate-length orbital periods. However, 52\% of satellites have $D_2-D_1 > +15$ kpc. Thus, most satellites are approaching the MW today, while only 14\% are moving away from the MW ($D_2-D_1 > -15$ kpc). These trends explain the gradually changing gray differential in the bottom left panel of Figure \ref{fig:crd_main}, where the satellites moving towards the MW contribute to where $\rm \Delta N_{sat}$ is the smallest at $D=25-125$ kpc. In contrast, the smaller fraction of satellites moving away from the MW gives rise to the changes in the differential at $D > 125$ kpc. 

In contrast to the bottom left panel of Figure \ref{fig:crd_main}, the bottom right panel of Figure \ref{fig:crd_main} shows the magnitude of $\rm \Delta N_{sat}(<D)$ is relatively constant over time. Changes in the differential radial profiles are most significant between $D=25-75$ kpc. Between present-day and 1 Gyr ago, seven satellites with long period orbits move toward the MW from $D_1\sim75-150$ to $D_0=25-75$ kpc. This shift of satellites toward the inner MW halo corresponds to the changes in the black differential in the bottom right panel of Figure \ref{fig:crd_main} between $D=25-75$ kpc. 

These satellites on long-period orbits moving toward the MW also give rise to the sharp increase of satellites in the cumulative radial profiles at present-day distances of $D\sim25-75$ kpc in both bottom panels of Figure \ref{fig:crd_main}. Similarly, they also affect the evolution of $R_{50}$, as shown in Figure \ref{fig:r50_time}, such that $R_{50}$ steadily declines between 1 Gyr and today in the rigid MW+LMC potential and from 0.5 Gyr to present day for the rigid MW potential.

\bibliography{references}

\end{document}